\documentclass[useAMS,usenatbib]{mnras}
\usepackage{times,amsmath,amssymb,txfonts,graphicx,nicefrac,bm}
\usepackage{xspace,color}
\usepackage{color,ulem,soul}

\renewcommand\emph[1]{\textit{#1}}

\definecolor{dark-red}{rgb}{0.75, 0.00, 0.00}
\definecolor{hlcolor}{rgb}{1.00, 0.90, 0.85}\sethlcolor{hlcolor}

\newcommand{\Section}[1]{Section~\ref{sec:#1}}

\newcommand{\Fig}[1]{Fig.~\ref{fig:#1}}
\newcommand{\Figure}[1]{Figure~\ref{fig:#1}}

\newcommand{\Table}[1]{Table~\ref{tab:#1}}

\newcommand{\Equation}[1]{equation~(\ref{eq:#1})}
\newcommand{\Equations}[2]{equations~(\ref{eq:#1}) and (\ref{eq:#2})}

\newcommand\erg{\,\rm erg}

\newcommand\Myr{\,\rm Myr}

\newcommand\kms{\,\rm km\,s^{-1}}
\newcommand\pc{\,\rm pc}

\newcommand\kpc{\,\rm kpc}

\newcommand\tms{\!\times\!}
\newcommand\cdt{\!\cdot\!}

\newcommand\xx{\hat{{\mathbf x}}}
\newcommand\yy{\hat{{\mathbf y}}}
\newcommand\zz{\hat{{\mathbf z}}}

\newcommand\V{\mathbf v}
\newcommand\B{\mathbf B}

\newcommand\mB{\mn{\mathbf{B}}}

\newcommand\EMF{\mbox{\boldmath{${\cal E}$}}}
\newcommand\Tf{\mathcal{B}}
\newcommand\mTf{\overline{\mathcal{B}}}

\newcommand\tauc{\tau_{\rm c}}
\newcommand\etat{\eta_{\rm T}}

\newcommand\kf{k_{\rm f}}
\newcommand\kk{\bm{k}}

\newcommand{\mn}[1]{\overline{#1}}

\newcommand\Rm{\mathrm{Rm}}

\newcommand{\simgt}%
           {\,\hbox{\lower0.35ex\hbox{$\sim$}\llap{\raise0.35ex\hbox{$>$}}}\,}
\newcommand{\simlt}%
           {\,\hbox{\lower0.35ex\hbox{$\sim$}\llap{\raise0.35ex\hbox{$<$}}}\,}

\newcommand\NIRVANA{\textsc{nirvana}\xspace}
\newcommand\NIII{\textsc{nirvana-iii}\xspace}

\title[Non-local dynamo effects in ISM turbulence]%
      {On the spatial and temporal non-locality of dynamo mean-field effects
        in supersonic interstellar turbulence}
      \author[Gressel \& Elstner]%
             { Oliver~Gressel$^{1,2}$\thanks{E-mail:~ogressel@aip.de} and
               Detlef~Elstner$^1$\thanks{E-mail:~delstner@aip.de}\\
               $^1$Leibniz-Institut f{\"u}r Astrophysik Potsdam (AIP),
               An der Sternwarte 16, 14482, Potsdam, Germany\\
               $^2$Niels Bohr International Academy, The Niels Bohr Institute,
               Blegdamsvej 17, DK-2100, Copenhagen \O, Denmark\\
             }

\begin{document}

\date{Accepted 1988 December 15. %
      Received 1988 December 14; %
      in original form 1988 October 11}

\pagerange{\pageref{firstpage}--\pageref{lastpage}} \pubyear{2012}

\maketitle

\label{firstpage}


\begin{abstract}
  The interstellar medium of the Milky Way and nearby disk galaxies harbours large-scale coherent magnetic fields of Microgauss strength, that can be explained via the action of a mean-field dynamo.
  As in our previous work, we aim to quantify dynamo effects that are self-consistently emerging in realistic direct magnetohydrodynamic simulations, but we generalise our approach to the case of a non-local (non-instantaneous) closure relation, described by a convolution integral in space (time).
  To this end, we leverage our comprehensive simulation framework for the supernova-regulated turbulent multi-phase interstellar medium. By introducing spatially (temporally) modulated mean fields, we extend the previously used test-field method to the spectral realm -- providing the Fourier representation of the convolution kernels.
  The resulting spectra of the dynamo mean-field coefficients that we obtain broadly match expectations and allow to rigorously constrain the degree of scale separation in the Galactic dynamo. A surprising result is found for the diamagnetic pumping term, which \emph{increases} in amplitude when going to smaller scales.
  Our results amount to the most comprehensive description of dynamo mean-field effects in the Galactic context to date. Surveying the relevant parameter space and quenching behaviour, this will ultimately enable the development of assumption-free sub-grid prescriptions for otherwise unresolved global galaxy simulations.
\end{abstract}



\section{Introduction}
\label{sec:intro}

Large-scale dynamo action by turbulence, rotation and shear in the interstellar medium (ISM) is a main process for the explanation of coherent magnetic fields in galaxies \citep{1996ARA&A..34..155B,2010ASP..conf..197F,2015A&ARv..24....4B}. The dominant driver for the turbulence in the multi-phase ISM is the action of super-nova explosions \citep[e.g.,][]{1999ApJ...514L..99K,2004Ap&SS.292..207D,2006ApJ...653.1266J,2013MNRAS.430L..40G}. This scenario is in some sense comparable to models with forced turbulence, but with a forcing on a wider range of scales, starting on small scales by the explosions and injecting further energy on larger scales by the expanding shock waves \citep{2006MNRAS.370..415M}.  The turbulent ISM is extensively studied under various aspects for the observable interpretation. Our aim is a more detailed understanding of the magnetic field amplification.  We consider numerical simulations in a local box including rotation and shear, that lead to a large scale magnetic field growth by a dynamo \citep[as first demonstrated by][]{2008A&A...486L..35G}.  The properties of the turbulent electromotive force (EMF) are studied by the analysis of turbulent transport coefficients in several ways for various types of turbulent media.  Most prominent is the use of additional, passive test fields \citep{2009PhDT........99G}, but also other methods relying only on actual simulation data are applied.  A comparison of two methods for the turbulent ISM was given by \citet{2020MNRAS.491.3870B}.  Both methods gave a good approximation of the EMF by the mean magnetic field, but with slightly different values for the turbulent diffusivity.  We take now test fields with variations in space \citep*{2008A&A...482..739B} and time \citep{2009ApJ...706..712H} for the analysis of the turbulent EMF.  Unlike in earlier studies, we can investigate the action of turbulence on various length scales in the kinematic regime of the dynamo, and study the importance of so-called ``memory effects''.

This paper is organised as follows: In \Section{methods} we describe the equations of motion and introduce the mean-field closure relation along with a method of obtaining coefficients. In \Section{results} we then present results from a set of two turbulent ISM simulations. We discuss our results in \Section{discussion}.
\vspace{-5ex}


\section{Methods}
\label{sec:methods}

Using a modified version of the \NIII code \citep{2011JCoPh.230.1035Z}, we perform direct numerical simulations (DNS) of the magnetised, multi-phase turbulent interstellar medium, where the velocity dispersion is driven self-consistently by injection of supernova energy.

\subsection{Equations of motion and model parameters} 
\label{sec:model}

We solve the equations of viscous, non-ideal magnetohydrodynamics (MHD) in a local shearing-box frame of reference \citep{2007CoPhC.176..652G}, where we adopt an angular frequency of ${\Omega=4\times 25\kms\kpc^{-1}}$. Differential rotation is expressed via the parameter $q\equiv{\rm d}\ln\Omega/{\rm d}\ln R = -1$ for the Galactic rotation curve. As in previous work \citep{2008A&A...486L..35G,2013A&A...560A..93G,2015AN....336..991B}, we adopt a domain size of $(0.8\kpc)^2 \times \pm2.13 \kpc$, but with a moderately increased linear resolution of $\Delta=6.7\pc$ (instead of $\Delta=8.3\pc$). In Cartesian coordinates, ($x$, $y$, $z$), the equations read
\begin{eqnarray}
      \frac{\partial\rho}{\partial t}
      + \nabla\cdot\left(\rho \V\right) & = & 0\,,
      \nonumber\\
      \frac{\partial\left(\rho\V\right)}{\partial t}
      + \nabla\cdot\left(\rho\mathbf{vv}
      + \mathrm{p}^{\star}
      - \mathbf{BB}\right) & = &
      - 2 \rho\ \Omega\ \hat{z}\times\V
      + 2\rho\ \Omega^{2} q x\ \hat{x} \nonumber\\ & &
      + \rho\, \mathrm{g}(z)\ \hat{z} +\nabla\cdot\tau\,,
      \nonumber\\
      \frac{\partial e}{\partial t}
      + \nabla\cdot\left((e+\mathrm{p}^{\star})\,\V-
      \left(\V\!\cdot\!\mathbf{B}\right)\mathbf{B}\right) & = &
      + 2 \rho\ \Omega^{2} q x\ \hat{x}\cdot\V
      + \rho\, \mathrm{g}(z)\ \hat{z}\cdot\V
      \nonumber\\ & &
      + \nabla\cdot\left[\tau \V
      + \eta_{\rm m}\,\mathbf{B}\times\left(\nabla\times\mathbf{B}\right)\right]
      \nonumber\\ & &
      + \nabla\cdot(\kappa \nabla \mathrm{T})
      - \rho^2 \Lambda\left(\mathrm{T}\right)\nonumber\\ & &
      + \Gamma_{\rm SN} + \rho\ \Gamma_{\rm UV}(z)\,,
      \nonumber\\
      \frac{\partial\mathbf{B}}{\partial t}
      - \nabla\times\left(\V\times\mathbf{B} -\eta_{\rm m}
      \nabla\times\mathbf{B}\right) & = & 0\,,
      \label{eq:mhd}
\end{eqnarray}
where $\mathrm{p}^{\star}\equiv \mathrm{p}+\mathbf{B}^2/2$. We assume an ideal-gas law, $\mathrm{p} = \left(\gamma-1\right)\ \epsilon$ (with $\gamma=5/3$, the ratio of specific heats) and obtain the internal energy as $\epsilon=e - \rho\V^2/2 - \B^2/2$. We include the vertical gravity force, $g(z)$, from the stellar disk according to \citet{1989MNRAS.239..605K}, that is, $g(z) = -a_1\ z\ (z^2-z_0^2)^{-1/2}-a_2\ z$, with empiric constants $a_{1}=1.42\times10^{-3}\kpc \Myr^{-2}$, $a_{2}=5.49\times10^{-4}\Myr^{-2}$ and $z_{0}=180\pc$. Horizontal boundary conditions are shear-periodic, and we apply standard outflow conditions in the vertical direction.

We include a small heat conduction term, $\nabla\cdot(\kappa \nabla \mathrm{T})$ with conduction coefficient $\kappa$, to stabilise the thermal instability \citep{1965ApJ...142..531F} near the grid scale \citep[see][]{2009A&A...498..661G}. The viscous stress tensor has the usual form of $\tau \equiv \tilde{\nu}_{\rm m}\,[\nabla\V + (\nabla\V)^{\top} - \nicefrac{2}{3} \, (\nabla\cdot\V)]$. With values $\tilde{\nu}_{\rm m}=0.5 \times 10^{25}\,{\rm cm^2s^{-1}}$ and $\eta_{\rm m}= 2\times 10^{24}\,{\rm cm^2s^{-1}}$, we obtain a magnetic Prandtl number of ${\rm Pm}\equiv \nu_{\rm m}/\eta_{\rm m}=5/2$. The various energy source and sink terms \citep[see][for details]{2008AN....329..619G,2015AN....336..991B} include an interstellar cooling function, $\Lambda(T)$, heating from the ambient stellar UV field, $\Gamma_{\rm UV}(z)$, as well as localised injection of SN energy, $\Gamma_{\rm SN}$. Galactic supernova rates are taken as $\sigma_{\rm I} = 4\Myr^{-1}\kpc^{-2}$ and $\sigma_{\rm II} = 30\Myr^{-1}\kpc ^{-2}$, for type~I$\,$/$\,$II SNe, respectively. The energies associated with the SNe are $E_{\rm I}=10^{51}$ and $E_{\rm II}=1.14 \times10^{51}\erg$, and we crudely mimic the effect of massive star associations via spatial clustering for type~II events based on a simple density threshold criterion \citep{1999ApJ...514L..99K}.

\subsection{Local versus non-local closure relations} 

The heart of the mean-field approach to modelling dynamo activity in a turbulent plasma is the mean (or sometimes called ``turbulent'') electromotive force, $\EMF \equiv \overline{\V'\tms \B'}$, where the bar denotes averaging over horizontal slabs, and $\B'\equiv\B-\mB$, and $\V'\equiv\V-\mn{\V}$. The EMF exhaustively captures the effects of \emph{correlated}  velocity and magnetic field fluctuations. It appears as an additional source in the mean-field induction equation
\begin{equation}
    \partial_t \mB - \nabla\tms(\mn{\V}\tms\mB) = \nabla\tms\EMF\,,
  \label{eq:MF_ind}
\end{equation}
describing the evolution of the mean magnetic field. The EMF is typically expressed as a linear functional of the mean magnetic field and its gradients,\footnote{In the local frame, the radial and azimuthal directions are assumed to be homogeneous, which limits us to considering vertical gradients.} that is,
\begin{equation}
  \EMF_i(z,t) = \alpha_{i\!j}(z,t)\ \mn{B}_j(z,t)
          \ -\ \eta_{i\!j}(z,t)\ \varepsilon_{\!jzl}\,\partial_z \mn{B}_l(z,t)\,.
          \label{eq:closure}
\end{equation}
Here, the steady-state statistical properties of the chaotic flow are moulded into a pair of second-rank tensors, $\alpha_{i\!j}(z,t)$ and $\eta_{i\!j}(z,t)$. Note that these closure coefficients \emph{locally} and \emph{instantaneously} relate $\EMF(z,t)$ to the mean magnetic field, $\mn{B}(z,t)$, and its curl $\varepsilon_{\!jzl}\,\partial_z \mn{B}_l(z,t)$, expressed via the Levi-Civita tensor, $\varepsilon$. This local relation can be mathematically formalised by postulating a ``scale separation'' between $\mn{\B}$ and $\B'$ \citep[see][]{1980mfmd.book.....K}, pertaining both spatial and temporal variations.

Conversely to this local interpretation of the closure, the multi-scale character of turbulence suggests that the domain of dependence of $\EMF(z,t)$ is finite. This entails both non-local dependencies in space, as well as so-called ``memory effects'' \citep[e.g.,][]{2009ApJ...706..712H}.

\subsubsection{Scale-dependent mean-field effects} 

Accounting for this generalisation requires to replace the simple multiplicative relation in \Equation{closure} by a convolution operation \citep[e.g.,][]{2008A&A...482..739B}. In this case, the tensor coefficients, $\alpha$ and $\eta$, take the role of integration kernels. As for the temporal non-locality, this implies
\begin{equation}
  \EMF_i(z) = \int \hat{\alpha}_{i\!j}(z,\zeta)\ \mn{B}_j(z-\zeta)
                \ -\ \hat{\eta}_{i\!j}(z,\zeta)\ \varepsilon_{\!jzl}\,\partial_z
                \mn{B}_l(z-\zeta)\; {\rm d}\zeta\,.
  \label{eq:closure_conv_space}
\end{equation}
Note that $\hat{\alpha}_{i\!j}(\zeta)$ and $\hat{\eta}_{i\!j}(\zeta)$ may, in fact, still depend on $z$, reflecting the (slowly) changing conditions with height. Also note that \Equation{closure} is recovered in the limit when $\hat{\alpha}_{i\!j}(z,\zeta) \!=\! \alpha_{i\!j}(z)\, \delta(\zeta)$ and $\hat{\eta}_{i\!j}(z,\zeta) = \eta_{i\!j}(z)\, \delta(\zeta)$ in \Equation{closure_conv_space}, where $\delta(\zeta)$ is Dirac's delta distribution, corresponding to a flat spectrum in k-space. Omitting the $z$~dependence, \Equation{closure_conv_space} in Fourier space simply becomes
\begin{equation}
  \tilde{\EMF}_i(k_z) = \tilde{\alpha}_{i\!j}(k_z)\ \tilde{\mn{B}_{\!j}}(k_z) \ -\ \tilde{\eta}_{i\!j}(k_z)\ {\rm i} k_z \,\varepsilon_{\!jzl} \,\tilde{\mn{B}_l}(k_z) \,,
  \label{eq:closure_fourier_space}
\end{equation}
which defines the context for scale-dependent closure coefficients. Equations~(\ref{eq:closure_conv_space}) and (\ref{eq:closure_fourier_space}) illustrate the correspondence between scale dependence, on the one hand, and non-locality, on the other hand. Studying isotropic turbulence, \citet{2008A&A...482..739B} have found coefficients with an approximate Lorentzian dependence
\begin{equation}
  \tilde{\alpha}(k_z) =
   \frac{\alpha_{0}}{1+\left(k_z/k_{\rm c}^{(\alpha)}\right)^2}\,,\qquad
  \tilde{\eta}(k_z)   =
   \frac{\eta_{0}}  {1+\left(k_z/k_{\rm c}^{(\eta)} \right)^2}\,,
  \label{eq:kernel_fourier}
\end{equation}
in Fourier space, with characteristic scales $k_{\rm c}^{(\alpha)}$, and $k_{\rm c}^{(\eta)}$ that may indeed be different for the two types of effects. Note that \Equation{kernel_fourier} implies an exponentially decaying convolution kernel, $\hat{\alpha}(\zeta)\propto {\rm exp} -k_{\rm c}\zeta\;$ in real space. Such a dependence was also found for helical driving with shear \citep{2009A&A...495....1M}, for passive scalar diffusion \citep{2010PhRvE..82a6304M} and, more recently, for the case of magnetorotational turbulence \citep{2015ApJ...810...59G}.

\subsubsection{Non-instantaneous mean-field response} 

In the same way, as \Equation{closure_conv_space} describes the non-locality of the closure relation in space, one can postulate that the EMF is a non-instantaneous response to the presence of a mean magnetic field -- and therefore depends on its history. Formally, this approach is related to the so-called $\tau$~approximation, which extends the second-order correlation\footnote{This is sometimes also referred to as ``first-order smoothing'' (FOSA).} approximation (SOCA). The $\tau$~approximation provides a closure relation by replacing third-order expressions with second-order ones multiplied by a characteristic time, $\tau$.

In analogy to \Equation{closure_conv_space}, the non-instantaneous closure relation is formulated as a convolution integral in time, that is
\begin{equation}
  \EMF_i(z,t) = \int
  \hat{\alpha}_{i\!j}(z,t')\,\mn{B}_j(z,t-t') \ -\
  \hat{\eta}_{i\!j}(z,t')\ \varepsilon_{\!jzl}\,\partial_z\mn{B}_l(z,t-t')
  \ {\rm d}t'\,,
  \label{eq:closure_conv_time}
\end{equation}
where, in our geometry, the integral kernels $\hat{\alpha}_{i\!j}(t')$ and $\hat{\eta}_{i\!j}(t')$ again remain functions of the vertical coordinate, $z$. Cast into Fourier space, this relation becomes multiplicative, that is
\begin{equation}
  \tilde{\EMF}_i(\omega) = \tilde{\alpha}_{i\!j}(\omega)\ \tilde{\mn{B}_{\!j}}(\omega) \ -\ \tilde{\eta}_{i\!j}(\omega)\ {\rm i} k_z \,\varepsilon_{\!jzl} \,\tilde{\mn{B}_l}(\omega) \,.
  \label{eq:closure_fourier_time}
\end{equation}
For this closure relation, \citet{2009ApJ...706..712H} have demonstrated that, to a reasonable degree of approximation, one can model the frequency-dependence as an ``oscillating decay'' of the form $\propto \Theta(t)\,{\rm e}^{-t/\tauc}\cos(\omega_0 t)$, where $\Theta(t)$ denotes the Heaviside step function. When translated to Fourier space, this implies spectral dependencies of the form
\begin{equation}
  \tilde{\alpha}(\omega) = \alpha_0\;
  \frac{ 1-{\rm i}\,\omega\,\tauc^{(\alpha)}}%
       { \left(1-{\rm i}\, \omega\,\tauc^{(\alpha)}\right)^2 +
         \left(\omega_0^{(\alpha)}\,\tauc^{(\alpha)}\right)^2 }\,,
  \label{eq:osc_decay}
\end{equation}
with coefficients $\alpha_0$, $\tauc^{(\alpha)}$, and $\omega_0^{(\alpha)}$. A corresponding spectral expression can be written down for $\tilde{\eta}(\omega)$, with coefficients replaced by $\eta_0$, $\tauc^{(\eta)}$, and $\omega_0^{(\eta)}$, respectively.

In the present work, we aim to obtain scale-dependent closure coefficients in their form appearing in \Equations{closure_fourier_space}{closure_fourier_time}. These naturally lend themselves to determination via the test-field (TF) method \citep{2005AN....326..245S,2007GApFD.101...81S} in its spectral flavour. We follow the approach introduced by \citet*{2008A&A...482..739B} for \Equation{closure_fourier_space} and \citet{2009ApJ...706..712H} for \Equation{closure_fourier_time}, respectively. While a generalised approach combining \Equation{closure_conv_space} and (\ref{eq:closure_conv_time}) is indeed possible \citep[see][]{2012AN....333...71R}, we here focus on the two limiting cases and only assess memory effects at the largest scale (i.e., at $k_z^{\rm TF}\!=\!2\pi/L_z$), as well as the non-locality for the instantaneous response (i.e., at $\omega^{\rm TF}\!=\!0$).

\subsection{The spectral test-field method} 

The power of the TF method  stems from its well-behaved, externally imposed ``test fields'', $\mTf_{(\nu)}(z)$, which replace the potentially degenerate mean field, $\mn{\B}(z,t)$, that develops in the simulation itself. Chosen correctly, the TFs span a non-degenerate basis for determining all tensor coefficients in an unambiguous manner. To invert \Equation{closure_fourier_space} and (\ref{eq:closure_fourier_time}), respectively, and solve for the closure coefficients in a spectral sense, we apply the flavour of the method where the TFs are quadruplets of trigonometric functions\footnote{See \citet{2005AN....326..787B}. Note that formally, unlike for \citet{2005AN....326..245S}, who used polynomials, higher-order derivatives are non-zero here.}
\begin{eqnarray}
  \mn{\mathcal B}_{(0)} = \cos(\omega t)\, \cos(k_z z)\,\xx\,, &\quad
  \mn{\mathcal B}_{(1)} = \cos(\omega t)\, \sin(k_z z)\,\xx\,, &\nonumber\\
  \mn{\mathcal B}_{(2)} = \cos(\omega t)\, \cos(k_z z)\,\yy\,, &\quad
  \mn{\mathcal B}_{(3)} = \cos(\omega t)\, \sin(k_z z)\,\yy\,. &
\end{eqnarray}
In the first simulation, we set $\omega=\omega^{\rm TF}=0$, and use five spectral modes $k_z=k_z^{\rm TF} = 1,2,4,8,16\times 2\pi/L_z$, where $L_z$ is the vertical extent of the box. In a second, independent DNS, we set $k_z^{\rm TF}=2\pi/L_z$ and use eleven spectral modes $\omega=\omega^{\rm TF} = \nicefrac{1}{32}, \nicefrac{1}{16}, \nicefrac{1}{8}, \dots, 16, 32 \times 2\pi/P_0$, with $P_0=2.5\Myr$. This means that we are solving 20 and 44 additional induction equations, respectively. In practice, the $\nu^{\rm th}$ TF fluctuation, $\Tf'(\mathbf{r},t)$, is integrated alongside the DNS as
\begin{equation}
  \partial_t\Tf' = \nabla \times \left[
      \V'\tms\mTf + \mn{\V}\tms\Tf' - \!\overline{\V'\tms\Tf'} + \V'\tms\Tf'
    - \!\eta_{\rm m}\!\nabla\tms\Tf' \right] \,,\label{eq:testfields}
\end{equation}
with the velocity $\V(\mathbf{r},t)$ taken from the DNS providing the only link to the ``real'' physical evolution. For each of the quadruplets of TFs independently, we then evaluate the corresponding mean electromotive force $\EMF^{(\nu)}\equiv\overline{\V' \times \Tf'_{(\nu)}}$. A solution to \Equation{closure_fourier_space}/(\ref{eq:closure_fourier_time}) in terms of the TFs is then given by
\begin{equation}
  \left(\begin{array}{c}
      \tilde{\alpha}_{i\!j}(k_z,\omega)\\[2pt]
      k_z\ \tilde{\beta}_{i\!jz}(k_z,\omega)
    \end{array}\right) = {\rm e}^{\rm{i}\omega t}
  \left(\begin{array}{cc}
      \,\cos(k_z z) & \sin(k_z z) \\[2pt] \!-\sin(k_z z) & \cos(k_z z)
  \end{array}\right)
  \left(\begin{array}{c}
      \EMF_i^{(2j\!-\!2)} \\[2pt]
      \EMF_i^{(2j\!-\!1)}
    \end{array}\right)\,,\label{eq:tf_solution}
\end{equation}
where the tensors $\tilde{\eta}_{i\!j}(k_z,\omega)$ and $\tilde{\beta}_{i\!jz}(k_z,\omega)$ are simply related as $\tilde{\eta}_{xx} = \tilde{\beta}_{xyz}$, $\ \tilde{\eta}_{xy} = -\tilde{\beta}_{xxz}$, $\ \tilde{\eta}_{yx} = \tilde{\beta}_{yyz}$, and $\ \tilde{\eta}_{yy} = -\tilde{\beta}_{yxz}$ \citep[see][]{2002GApFD..96..319B}. Assuming steady-state (and/or homogeneous) conditions, the RHS of \Equation{tf_solution} is typically accumulated in time (and/or space) to obtain a sufficient statistical basis, but the method can in principle even deliver time-varying $\alpha$ and $\eta$. Note that for $\omega^{\rm TF}\!=\!0$, we have ${\rm e}^{\rm{i}\omega t}=1$, yielding real coefficients, whereas complex coefficients arise in the time-dependent case --- reflecting a (frequency-dependent) phase shift of the resulting EMF with respect to the causing mean-field.

In contrast to least-square methods \citep[see, e.g., discussion in][]{2020MNRAS.491.3870B}, that rely on a given mean-field configuration of interest to be realised in the DNS, \Equation{tf_solution} can be directly and unambiguously computed for each imposed TF wavenumber/oscillation period, readily yielding spectral-dependent vertical profiles for the dynamo parameters. We benchmark our spectral TF implementation for the simple test case of helical forcing in an unstratified cubic domain. Results obtained in the kinematic limit of the helical dynamo are presented in Appendix~\ref{sec:benchmark}.


\section{Results}
\label{sec:results}

We report results from the kinematic stage of two identical simulations described in detail in \Section{model}. All plots show time averages over the first half Gyr of the run. We begin our discussion with the spatial non-locality of non-oscillating mean fields.

\subsection{Non-local effects} 

\begin{figure}
  \center\includegraphics[width=0.9\columnwidth]{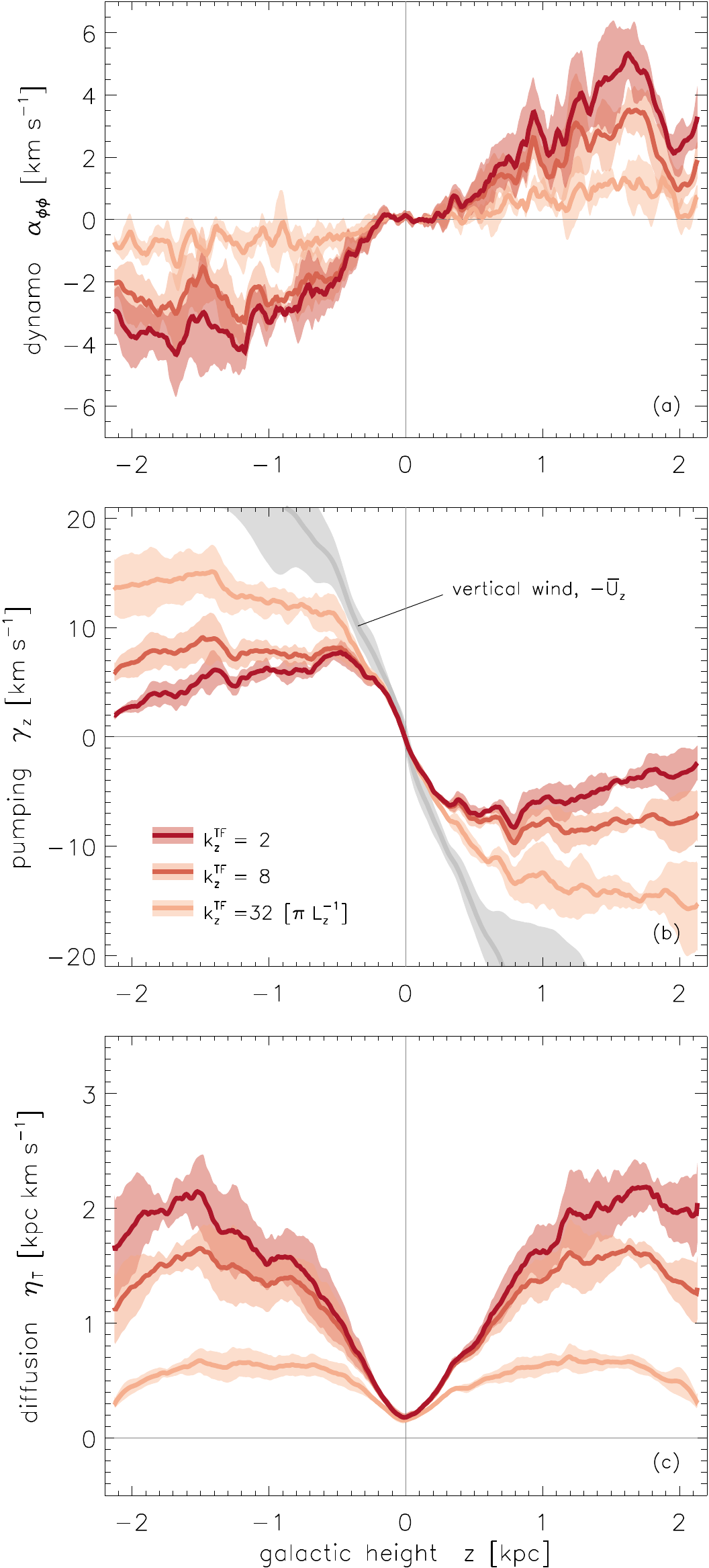}
  \caption{Vertical profiles of three fundamental mean-field dynamo coefficients in the kinematic regime. Panels show: a) azimuthal $\alpha$~effect, $\alpha_{\phi\phi}(z)$, b) vertical diamagnetic pumping, $\gamma_z(z)$, and c) vertical turbulent magnetic diffusion, $\etat(z)$. All coefficients are shown with $1\sigma$ fluctuations, and for three representative TF wavenumbers. Note the reverse trend (of higher amplitudes for increasing $k_z^{\rm TF}$) for $\gamma_z$ in the middle panel.}
  \label{fig:dyn}
\end{figure}

The three panels of \Figure{dyn} show vertical profiles of the three most central mean-field coefficients: a) the azimuthal $\alpha$~effect\footnote{We here use global cylindrical coordinates ($z$, $r$, $\phi$) to denote coefficients.}, $\alpha_{\phi\phi}(z)$, important for the $\alpha\Omega$~dynamo, b) the (downward) vertical diamagnetic pumping, $\gamma_z(z) \equiv \nicefrac{1}{2}(\alpha_{\phi r} - \alpha_{r\phi})$, and c) the turbulent diffusivity, $\etat(z) \equiv \nicefrac{1}{2}(\eta_{rr}+\eta_{\phi\phi})$. Results are shown for three (out of the five) TF wavenumbers as indicated by the key, and the middle panel, moreover, shows the outflow velocity $\mn{U}_z(z)$ (plotted with a minus sign) to illustrate its correspondence with $\gamma_z(z)$ within the inner $\pm 1\kpc$, which we attribute to the presence of a Galactic fountain in the clumpy multi-phase ISM.

While the $\alpha$~effect and turbulent diffusion show the expected trend towards diminished amplitudes when approaching smaller scales (i.e., higher TF wavenumber), the opposite is found for the vertical pumping term, $\gamma_z$. We speculate that this behaviour is related to the Galactic fountain flow, where the flux is anchored in the cold molecular phase, and where regions of converging flow --- further enhanced by the thermal instability --- collect the magnetic flux, producing a strong correlation between downward flow and mean horizontal magnetic field.

\begin{figure}
  \center\includegraphics[width=\columnwidth]{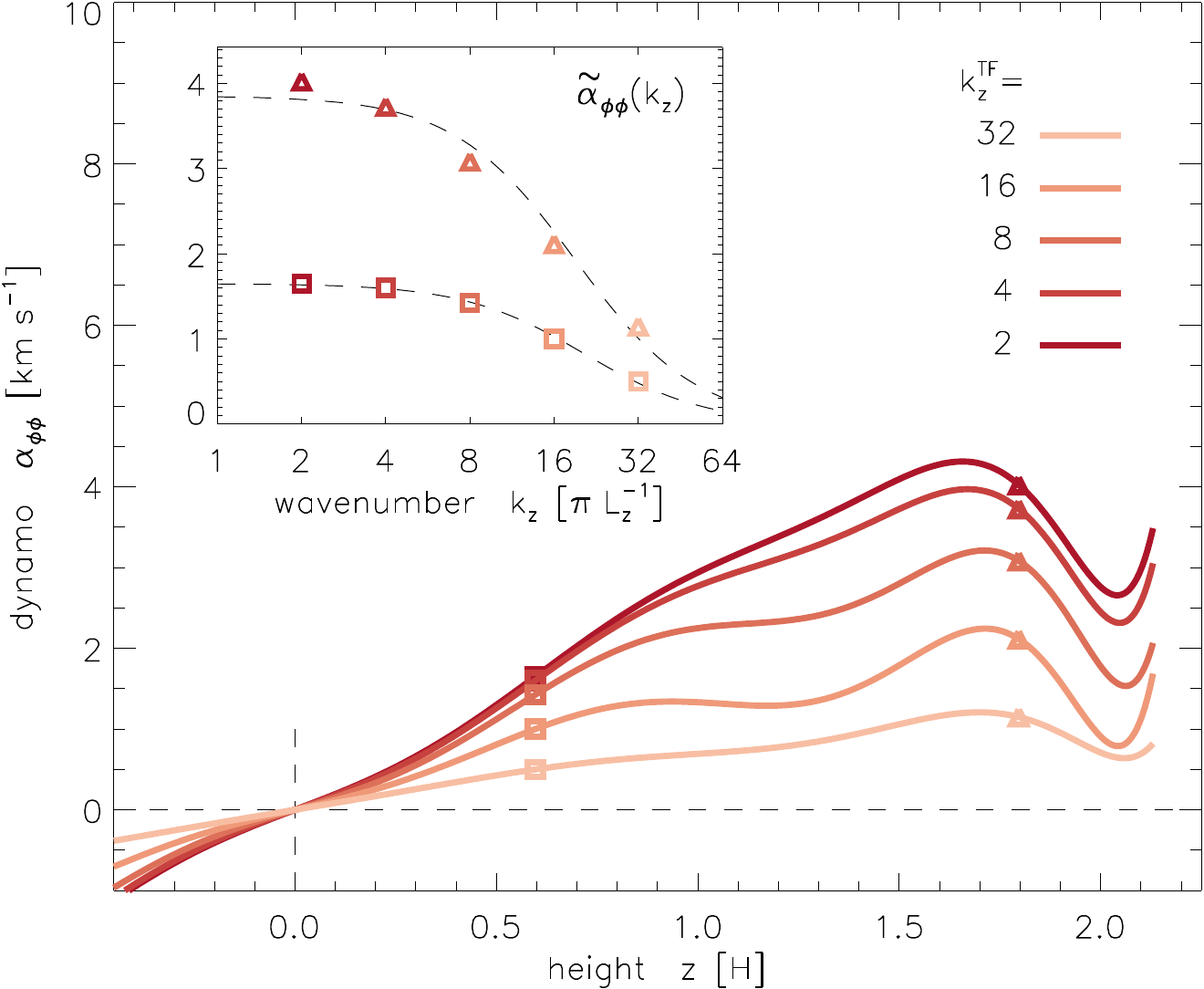}
  \caption{Spectral dependence of the azimuthal $\alpha$~effect, $\alpha_{\phi\phi}(z)$. The main plot shows the filtered vertical profiles for the five sampled TF wavenumbers, while the insets illustrates the spectral shape (fitted by Lorentzians, shown as dashed lines) for the two vertical locations $z=0.6$ and $1.8\kpc$, indicated by squares and triangles, respectively, in both plots.}
  \label{fig:k_dynamo}
\end{figure}



In \Figure{k_dynamo}, we attempt to quantify the spectral dependence of the profiles shown in \Fig{dyn}. To this end, we filter the obtained TF profiles by expanding them into a truncated series of Legendre polynomials up to order twelve, and sample point values (marked by `$\square$' / `$\triangle$') at locations $z=0.6$ and $1.8\kpc$, respectively. These sampled values are then shown in the inset of \Fig{k_dynamo} as a function of TF wavenumber, with dashed lines indicating a Lorentzian curve $\propto 1/(1+k_z^2/k_{\rm c}^2)$, with a characteristic scale $k_{\rm c}\simeq 20\pi/L_z$. Translated to real space, this corresponds to an exponential decay $\propto \exp(-k_{\rm c}\,|\zeta|)$ with a characteristic length $l_{\rm c} \equiv k_{\rm c}^{-1} \simeq 67\pc$, which is only moderately smaller than the inferred correlation length $\simeq 100\pc$ of the turbulent eddies in this type of simulation \citep[see sect.~3.4 of][]{2013ApJ...762..127B}.

\begin{figure}
  \center\includegraphics[width=\columnwidth]{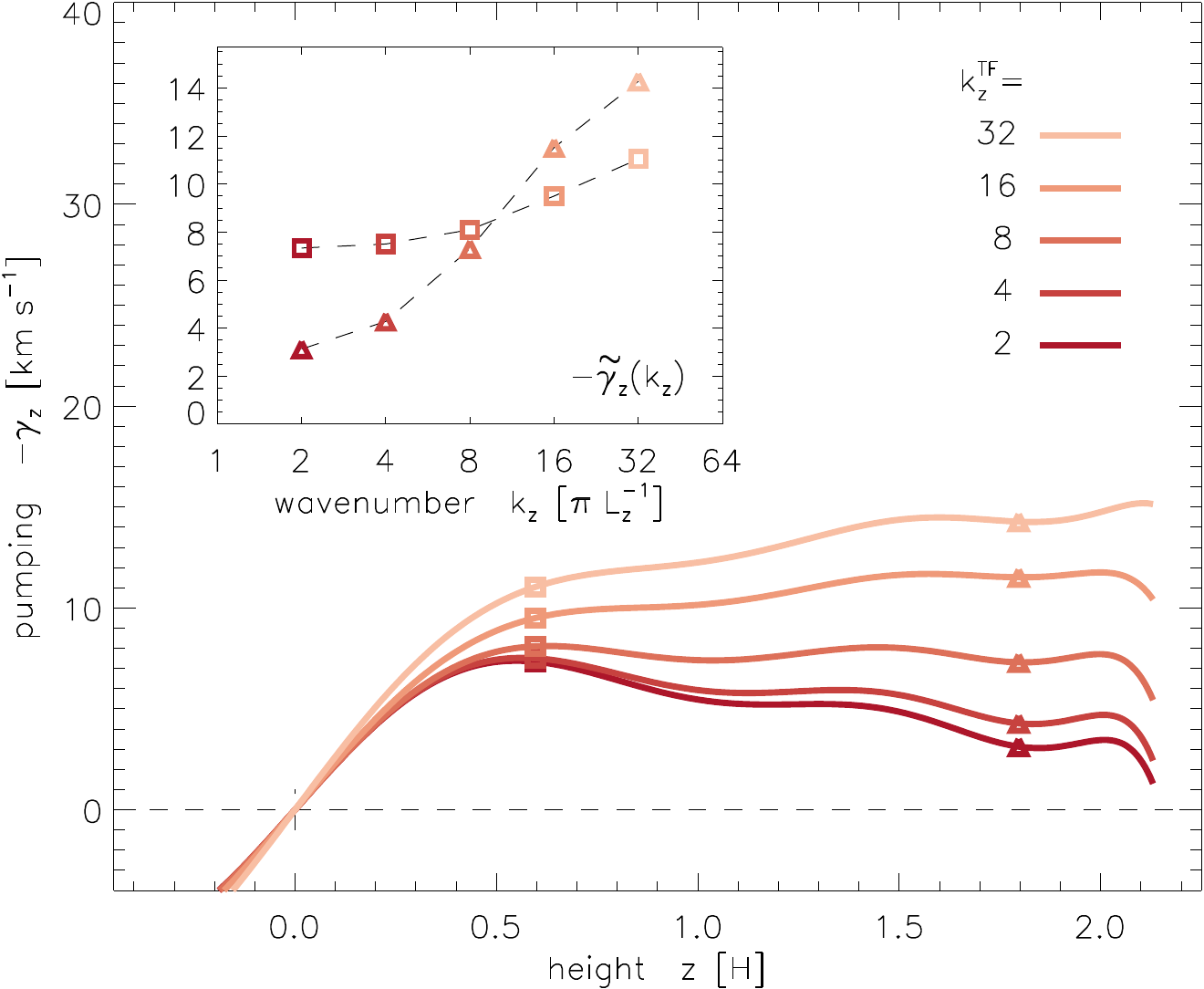}
  \caption{Same as \Fig{k_dynamo}, but for the vertical diamagnetic pumping, $\gamma_z(z)$. Note that, unlike for the other coefficients, the amplitude of the measured effect is boosted with increasing TF wavenumber, leading to a ``UV catastrophe''. No fit was attempted here, and dashed lines simply follow the data.}
  \label{fig:k_pumping}
\end{figure}

We follow the same protocol to extract the spectral dependence for the vertical pumping term, $\gamma_z$, which is shown in \Figure{k_pumping}, but without fitting a Lorentzian dependence to the curve. Contrary to expectations, we find the amplitude of the effect to \emph{increase} when going to smaller scales -- in particular at $z=1.8\kpc$, where $\gamma_z$ is boosted by a factor of five. Within $z\simlt 0.5\kpc$, one might argue that the amplitude remains approximately constant in k-space, corresponding to a $\delta$~function in real space, that is, a local relation between $\EMF(z)$ and $\mn{\B}(z)$ in terms of this contribution. In any case, this peculiar behaviour certainly warrants further investigation.

\begin{figure}
  \center\includegraphics[width=\columnwidth]{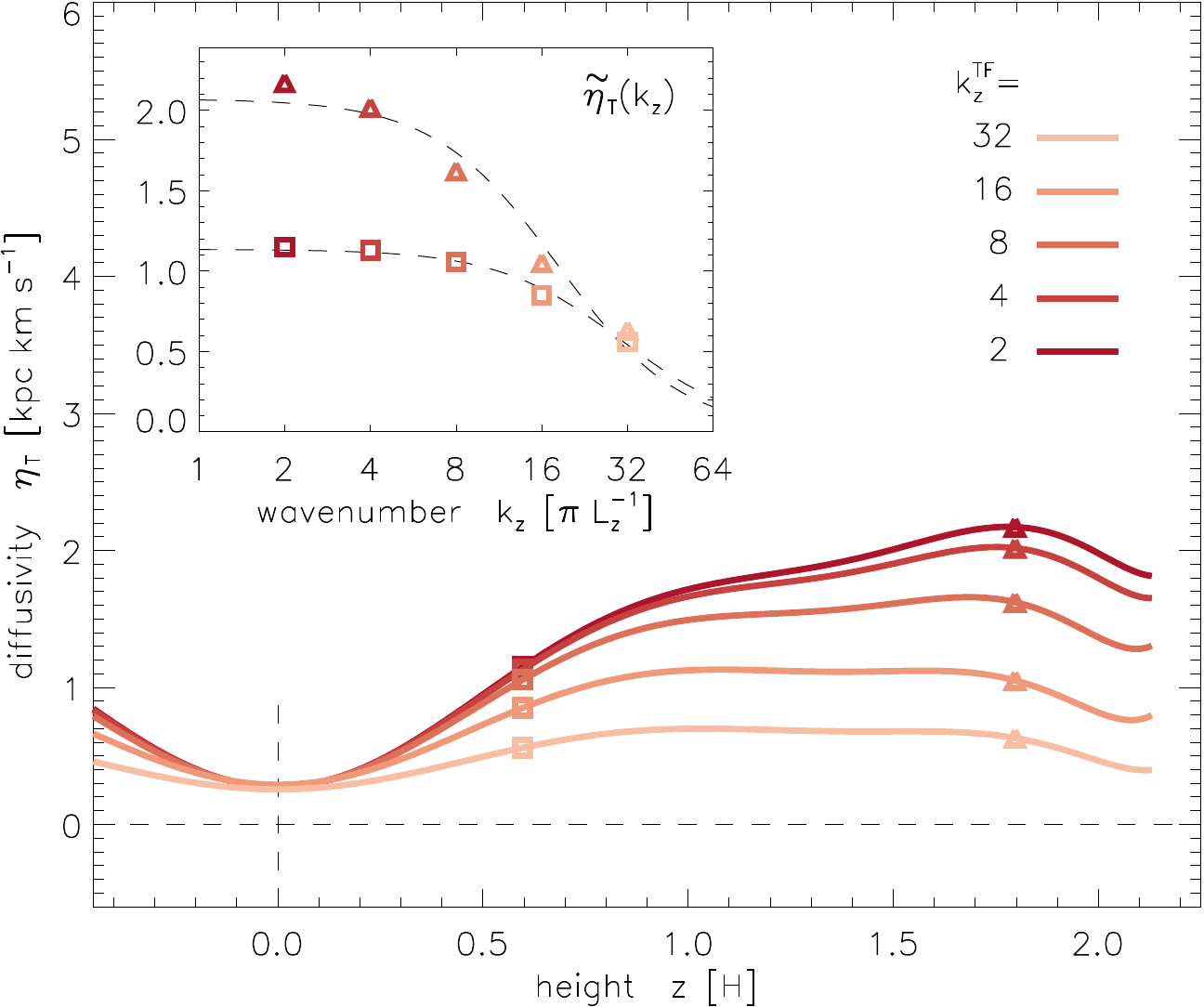}
  \caption{Same as \Fig{k_dynamo}, but for the vertical turbulent
    magnetic diffusion, $\etat(z)$. Dashed lines in the inset indicate Lorentzian curves fitted to the data, with a similar turn-over scale of $k_z^{\rm c}\simeq 16$ in units of $2\pi\,L_z^{-1}$.}
  \label{fig:k_diffusion}
\end{figure}

Finally, in \Figure{k_diffusion}, we show the spectral dependence of the turbulent diffusion, $\etat$, which again displays the regular behaviour of decaying amplitude in k-space. Unlike for $\alpha_{\phi\phi}$, there is a weak trend towards shorter correlation length $l_{\rm c}\simeq 45\pc$ at $z=0.6\kpc$ as compared to $z=1.8\kpc$. This is perfectly consistent with the inferred vertical dependence of the eddy size, which simply is a consequence of the strong density stratification within the ISM.

\begin{figure}
  \center\includegraphics[width=0.9\columnwidth]{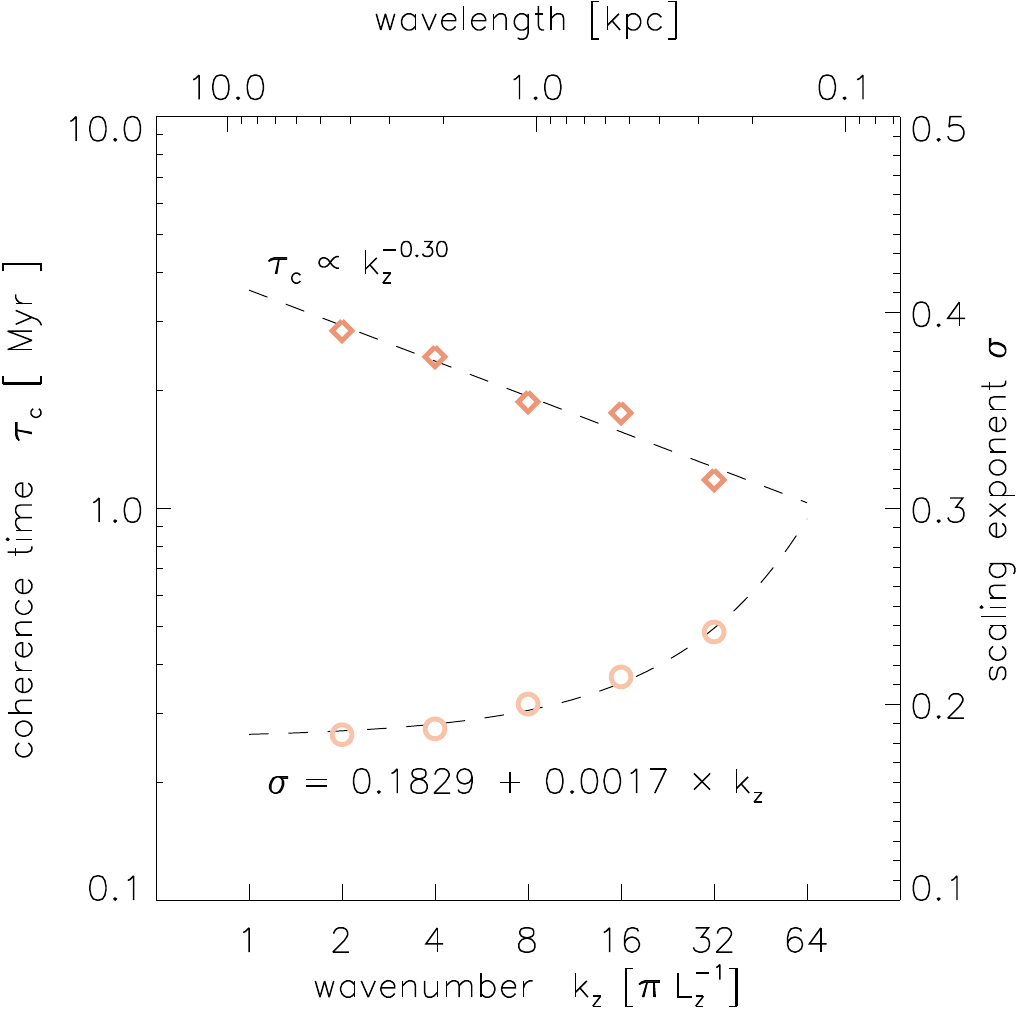}
  \caption{Scaling relation of the turbulent coherence time, $\tauc$ (`$\diamond$', left ordinate), and scaling exponent, $\sigma$ (`$\circ$', right ordinate), obtained via fitting the SOCA relation of \Equation{alpha_soca} to the TF coefficient $\alpha_{\phi\phi}(z)$.}
  \label{fig:k_tau_sigma}
\end{figure}

Following the approach described in detail in sect.~3.4 of \citet{2013ApJ...762..127B}, we fit the general SOCA expression
\begin{equation}
  \alpha_{\phi\phi}^{\rm soca}(z) =
  - \tauc^2\;\mn{u}_{\rm rms}^2(z)\; \Omega\,\zz\cdot\nabla\,\big[\ \mn{\rho}(z)^\sigma\; \mn{u}_{\rm rms}(z)\ \big]\,,
  \label{eq:alpha_soca}
\end{equation}
with free parameters ($\tauc$, $\sigma$) to our results for $\alpha_{\phi\phi}(z)$, using $\mn{\rho}(z)$ and $\mn{u}_{\rm rms}(z)$ as an input. The fit results are shown in \Figure{k_tau_sigma}, where we plot the obtained coefficients as a function of TF wavenumber. The $\sigma$ coefficient for compressible ISM turbulence had previously found to be $\sigma\simeq \nicefrac{1}{3}$ \citep{2013ApJ...762..127B}, whereas we here recover a somewhat smaller value of $\sigma\simeq 0.2$, with a weak scale dependence. Coherence times are $\simeq 3\Myr$ at the large scales, compatible with \citet{2013ApJ...762..127B} and are found to scale $\propto k_z^{-0.3}$. This is shallower than the typical $\propto k^{-2/3}$ scaling for the eddy turnover time in the Kolmogorov picture --- which is, however, not strictly applicable to the multi-phase ISM, that is highly compressible.

\subsection{Memory effects} 
\label{sec:memory}

\begin{figure}
  \center\includegraphics[width=\columnwidth]{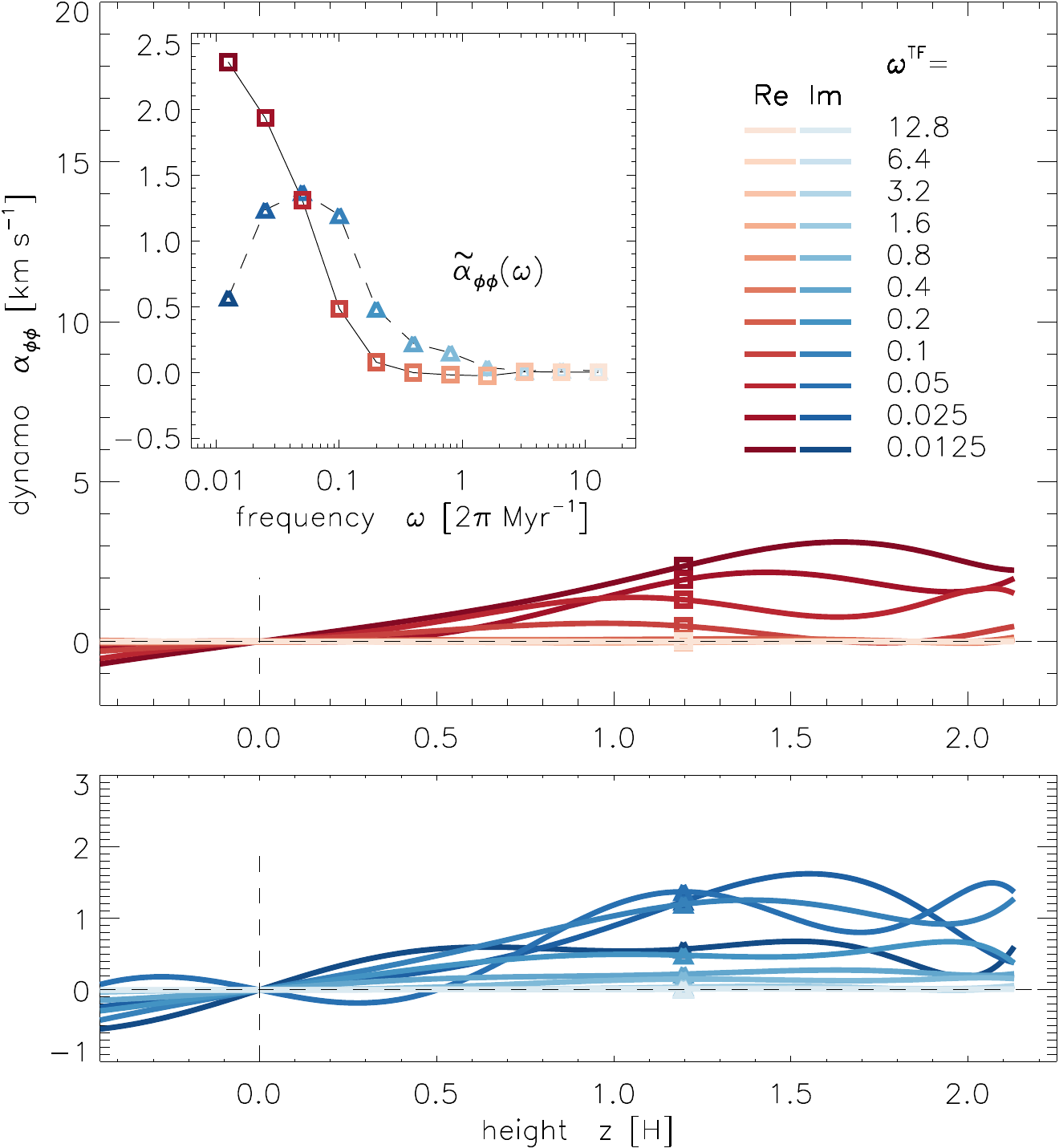}
  \caption{Real part (upper panel) and imaginary part (lower panel) of the frequency response of the dynamo $\alpha$ coefficient. The sampled solution at $z=1.2\kpc$ is plotted in the inset as a function of angular frequency of the pulsating TFs, spanning three decades in dynamic range.}
  \label{fig:w_dynamo}
\end{figure}

We now turn to the aspect of the EMF being a non-instantaneous response to an imposed (or, of course, self-consistently developing) mean-field. In \Figure{w_dynamo}, we plot the spectral dependence of the dynamo $\alpha$~effect as a function of oscillation frequency, $\omega^{\rm TF}$, covering three decades in dynamic range. As before, we spatially filter the vertical profiles to suppress by-chance fluctuations\footnote{Typically, these can be traced back to events including multiple SNe in a low-density region of the simulation volume, that leave a disproportionate imprint on the TFs and are of debatable relevance in the long-term mean.}. We then sample point values (here at $z=1.2\kpc$) to illustrate the spectral dependence. As such, the figure is only meant to convey the most basic picture in an accessible manner. Note that, opposed to previous figures, squares ('$\square$') now denote the real part of the coefficient, whereas triangles ('$\triangle$') represent the imaginary part. As probably expected for fully developed turbulence, the real part of $\alpha_{\phi\phi}(\omega)$ is restricted to frequencies below the (inverse) coherence time of the turbulence of $\tauc\simeq 3\Myr$. This means that more rapid oscillations in the mean fields presumably only accumulate uncorrelated $\V'\tms \B'$, averaging to negligible amplitude. A non-vanishing imaginary part at intermediate frequencies implies a certain phase lag between the mean-field and EMF. Since the $\alpha$~effect is ultimately caused by the Coriolis force, it would appear interesting to study this phenomenon as a function of Coriolis number, ${\rm Co}\equiv 2\,\Omega\tauc$, that is, the ratio of the Galactic rotation period and eddy turnover time. In our model with $\Omega=4\times$ the Galactic value of $25\kms\kpc^{-1}$, we have ${\rm Co}\simeq 0.6$, which should indeed put us in a transitional regime with respect to the importance of rotational effects.

\begin{figure}
  \center\includegraphics[width=\columnwidth]{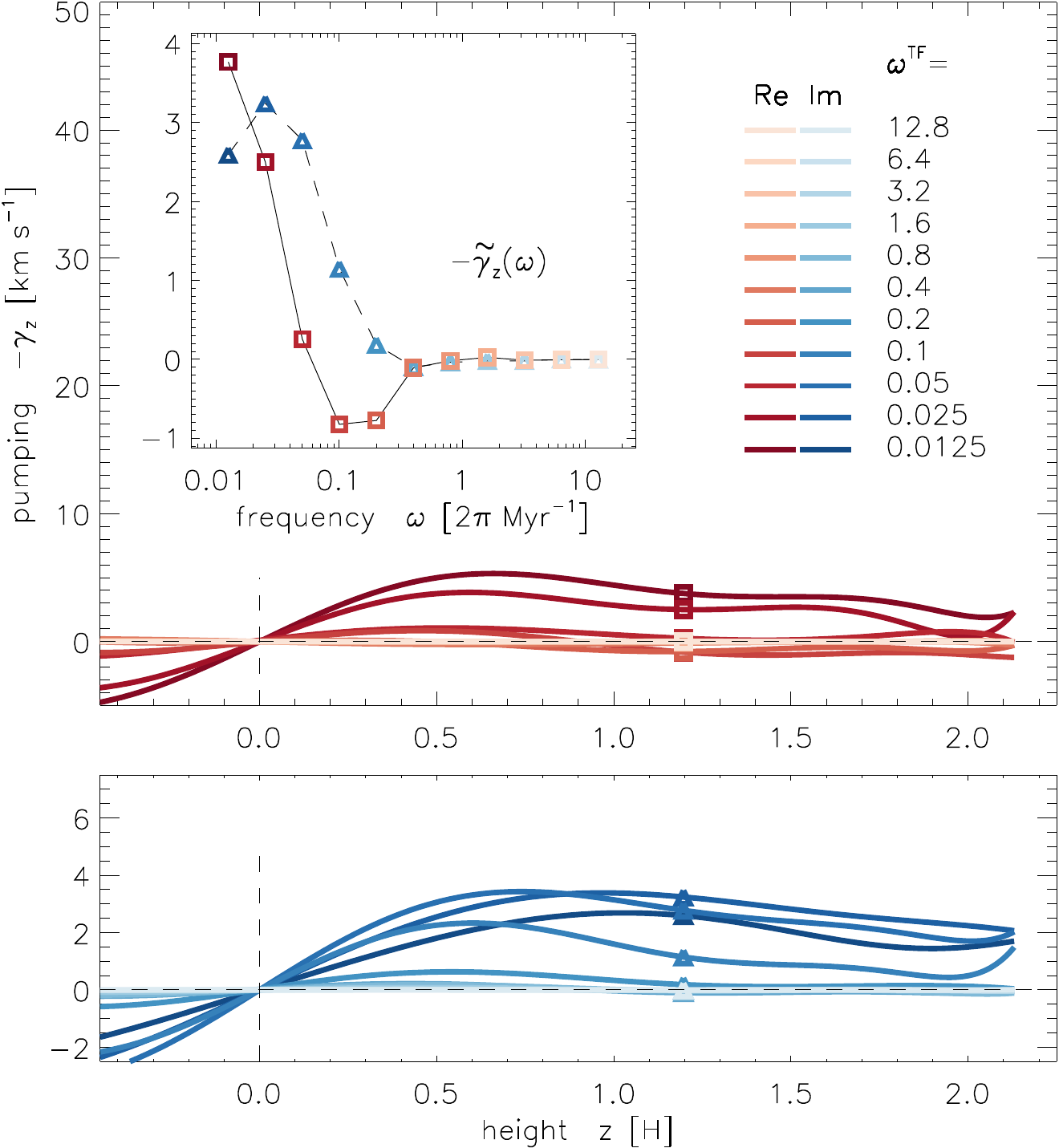}
  \caption{Same as \Fig{w_dynamo}, but for the vertical diamagnetic pumping, $\gamma_z(z)$. Note that the real part of $-\gamma_z(z)$ becomes negative at intermediate frequencies, potentially indicating some degree of \emph{upward} pumping.}
  \label{fig:w_pumping}
\end{figure}

The spectral response of the diamagnetic pumping term, $\gamma_z$ is plotted in \Figure{w_pumping}. In accordance with the unexpected scale-dependent behaviour seen in \Figure{k_pumping}, the real part of $-\gamma_z(\omega)$ as well shows non-standard behaviour in time, that is, it becomes negative at intermediate oscillation periods. At least in principle, this implies the possibility of turbulent advection of mean fields in the \emph{upward} direction. There, however, remains the ambiguity that, when translating the result into complex amplitude and phase, what appears as a sign change might in reality merely be a time-lag phenomenon. Unlike for $\alpha_{\phi\phi}(\omega)$, where both the real and imaginary part are positive, the phase lag here approaches $\pi/2$, however, which may appear questionable. Finally, as can be seen in the main plot of \Figure{w_pumping}, the real part of $-\gamma_z(\omega)$ is strictly positive for $z\simlt 0.6\kpc$.

\begin{figure}
  \center\includegraphics[width=\columnwidth]{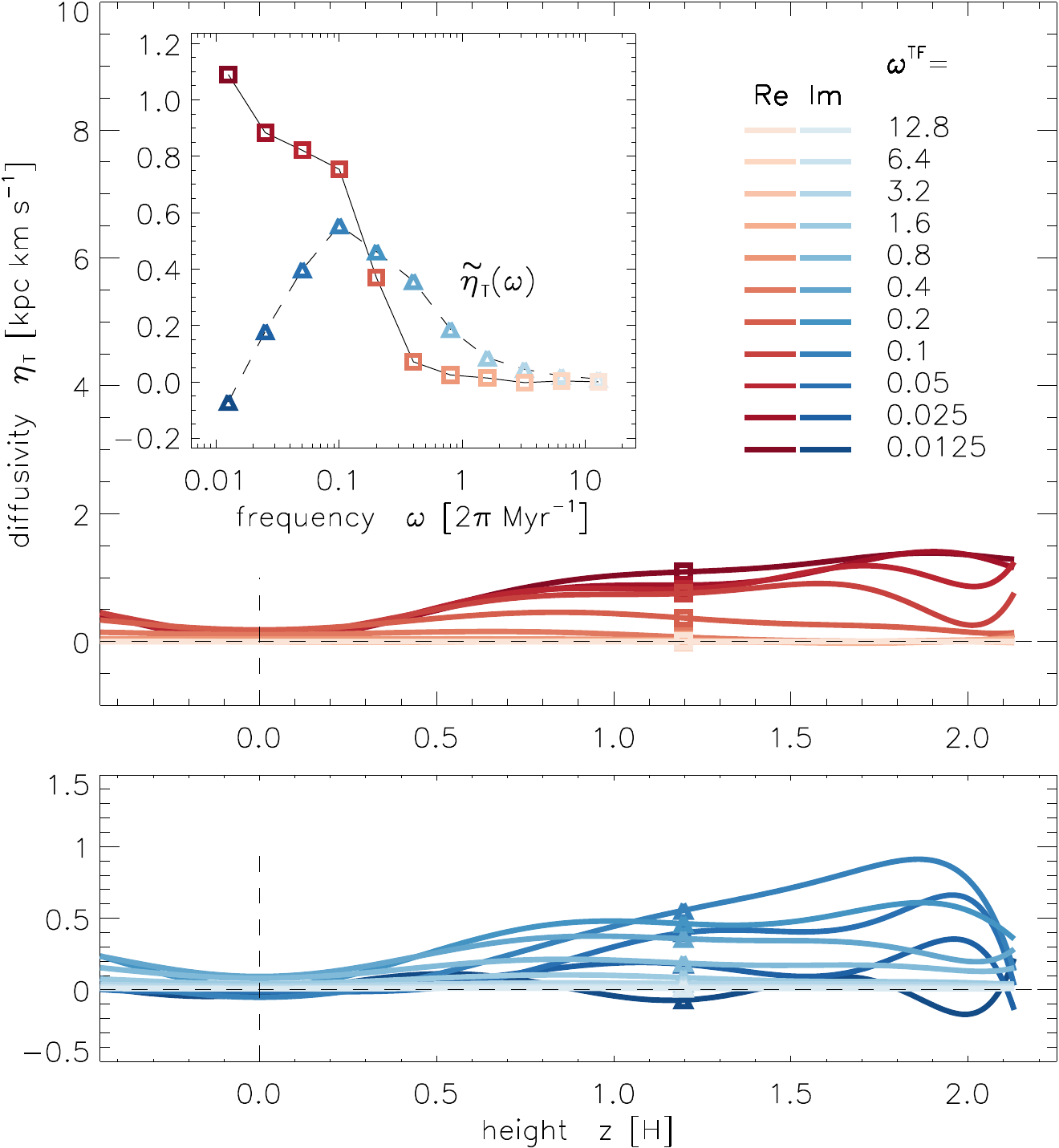}
  \caption{Same as \Fig{w_dynamo}, but for the turbulent magnetic diffusion, $\etat(z)$.}
  \label{fig:w_diffusion}
\end{figure}

In \Figure{w_diffusion}, we plot the spectral response of the turbulent magnetic diffusion, $\etat(\omega)$, which in general looks similar to \Figure{w_dynamo}. In both quantities, there is a trend from a dominant real part at low frequencies to a mixed state at intermediate oscillation frequencies, with the transition at slightly higher values in \Figure{w_diffusion}. In the range $0.3 < \omega < 3$, that is, for mean fields roughly matching the beat of the eddy turnover time, $\etat(\omega)$ displays a strong lag of up to $\pi$, i.e., fully out of phase.

\begin{figure}
  \center\includegraphics[width=\columnwidth]{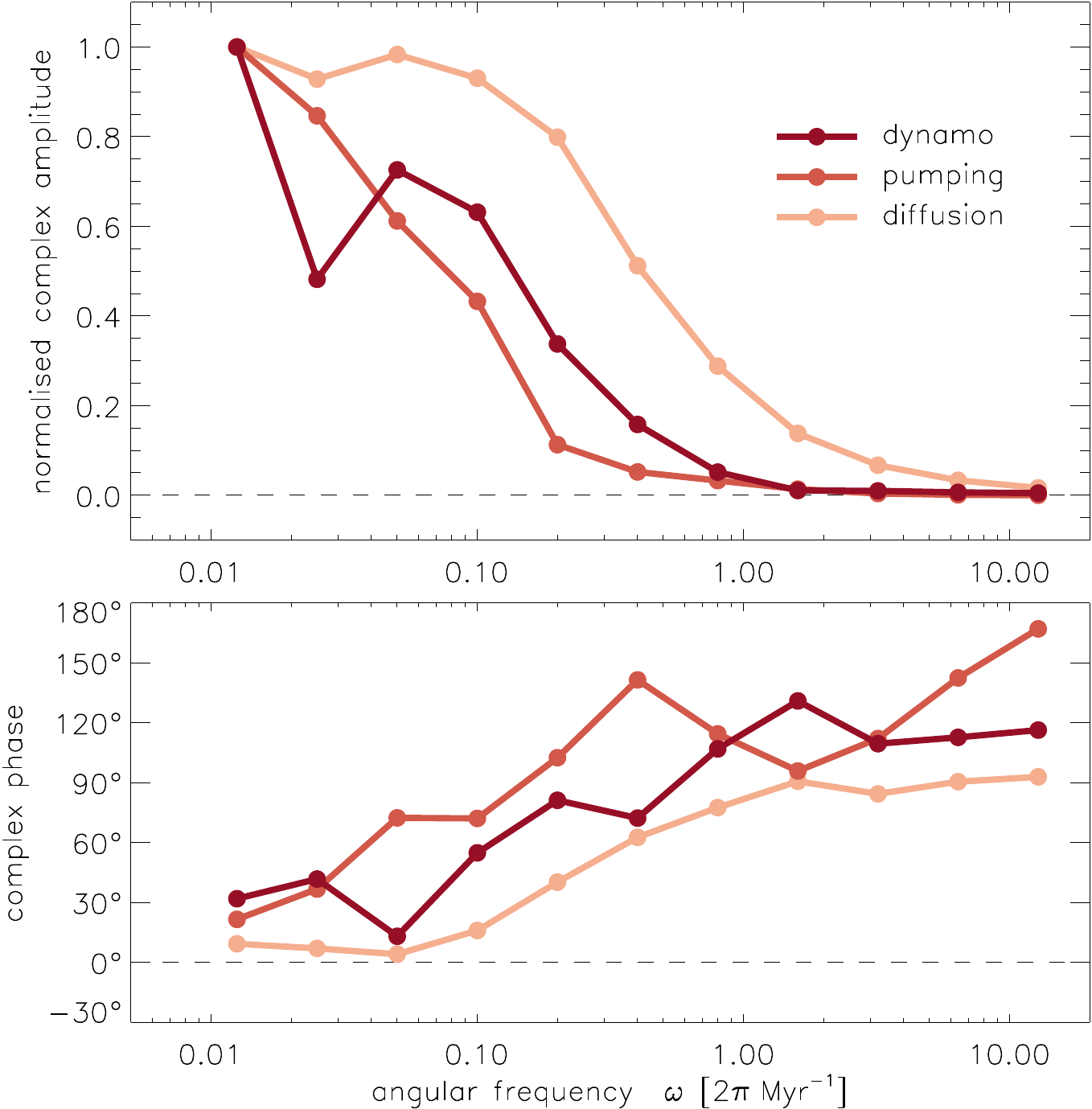}
  \caption{Normalised complex amplitude (upper panel) and phase (lower panel) illustrating the time-response of the three central mean-field effects.}
  \label{fig:w_coeff_cmp}
\end{figure}

Since in the discussion above we have already appealed to the notion of a phase lag, in \Figure{w_coeff_cmp}, we now plot the (normalised) complex amplitudes along with the complex phases for the three effects shown separately in the insets of Figures \ref{fig:w_dynamo}--\ref{fig:w_diffusion}, but now sampled at $z=0.6\kpc$. In terms of the amplitude of the effect as a function of frequency, it appears that the effects related to the $\alpha$~tensor share a common spectrum, whereas the diffusion term extends towards higher frequencies (by a factor of about three). This is probably not unexpected, but it would be interesting to see whether the width of this gap depends on certain input parameters such as, for instance, the Coriolis number. In the lower panel of \Figure{w_coeff_cmp}, it can be seen that the turbulent diffusion, $\etat$, displays a moderate phase lag between about $30^\circ$ and $90^\circ$, while that of the $\alpha$~effect as well as the pumping term, $-\gamma_z$, are somewhat larger at up to $150^\circ$.


\section{Discussion \& Conclusions}
\label{sec:discussion}

We have presented, for the first time, scale-dependent results of the mean-field closure coefficients found in self-consistent simulations of the turbulent interstellar medium. The reported profiles and scalings are derived with the TF method in the kinematic regime, where its validity is unchallenged.

\subsection{Scale dependence} 

For the component of the $\alpha$~effect, $\alpha_{\phi\phi}(z,k_z^{\rm TF}\!,\omega^{\rm TF}\!=\!0)$, that enters the classical $\alpha\Omega$~dynamo as well as for the turbulent diffusion, $\etat(z,k_z^{\rm TF}\!,\omega^{\rm TF}\!=\!0)$, we obtain a wavenumber dependence that is well approximated by a Lorentzian spectral dependence with a turnover at $k_{\rm c}^{-1}\simeq 67\pc$. This translates into a non-locality of the form $\hat{\alpha}(\zeta) \propto \exp(-k_{\rm c}\zeta)$, where $\zeta$ is the displacement in the vertical direction. How does this non-locality affect the evolution of the derived dynamo? To answer this question comprehensively, detailed mean-field calculations will be required \citep[see, e.g.,][who have demonstrated a change in the propagation of the dynamo wave for the case of MRI turbulence]{2002GApFD..96..319B}. In the following, we present a brief assessment in terms of dynamo numbers.

It is important to realise that the EMF enters the induction equation with a curl, that is, $\partial_t \mn{B}\propto \nabla\tms(\,\alpha\mn{B} - \eta\, \nabla\tms\mn{B}\,)$. Because products of the dynamo coefficients with the mean field appear inside the differential operation, one can distinguish two limiting cases. In the limit of uniform magnetic fields, the (in our case, vertical) gradients in the mean-field coefficients will matter. In this limit, the feasibility of the large-scale field amplification via the $\alpha\Omega$~mechanism can thus be gauged by looking at dimensionless dynamo numbers, $C_{\alpha} \equiv (\alpha/H_\alpha)\, (H_\eta^2/\eta_{\rm t})$ and $C_{\Omega} \equiv \Omega\, H_\eta^2/\eta_{\rm t}$, where $H_\alpha \equiv |\alpha|/|\partial_z\alpha| \simeq 0.45\kpc$ and $H_\eta \equiv |\etat|/|\partial_z\etat| \simeq 0.5\kpc$ are estimates for the integral scale of the dynamo-active region.\footnote{In fact, the profiles are well approximated by parabolas, making $H_\alpha(z)$ and $H_\eta(z)$ linear functions. The quoted values are at $z=0.5\kpc$.} We stress that this quantity is derived from the \emph{shape} of the profile shown in \Figure{k_dynamo}, which is in fact largely independent of $k_z^{\rm TF}$, and, as such, $H$ should not be confused with an inverse wavenumber, in this case. In fact, we obtain (comfortably super-critical) values of $C_\alpha C_\Omega\simeq 20$, completely independent of $k_z^{\rm TF}$.

In the converse limit, that is, for uniform amplitudes of the mean-field effects, but for spatially-varying mean fields, one can do a scale-by-scale estimation of the dynamo numbers. In this case, one of course has to replace $H_\alpha$ and $H_\eta$ by the wavelength of the Fourier mode under consideration. The resulting scaling of $C_\alpha C_\Omega \propto k^{-3}$ renders all but the fundamental modes stable to the mean-field dynamo instability, that is, $C_\alpha C_\omega < D_{\rm crit.}\simeq 10$. This explains why the $k_z=\pi/L_z$ (symmetric) and  $k_z=2\pi/L_z$ (anti-symmetric) modes dominate the evolution of the mean field in our simulations. At the same time, this justifies adopting the mean-field approach for SN-driven turbulence in the ISM.

As first demonstrated by \citet{2008A&A...486L..35G}, the \emph{downward} diamagnetic pumping caused by fountain flows plays a central role in enabling the Galactic dynamo. In that regard, the unusual divergent spectral behaviour of the $\gamma_z(z,k_z^{\rm TF}\!,\omega^{\rm TF}\!=\!0)$ term warrants further detailed investigation. Taken at face value, it would be interesting to study whether the precise spectral dependency is potentially related to the presence of Parker loops, and/or interchange-type instabilities. If so, they should likely also be found in simulations of cosmic-ray buoyancy-driven Parker instability \citep{2004ApJ...605L..33H,2006AN....327..469H,2007ApJ...668..110O,2019arXiv13588H}.

Fitting the SOCA relation to $\alpha_{\phi\phi}(z)$, using $\mn{\rho}(z)$ and $\mn{u}_{\rm rms}(z)$ as an input, we find $\tauc\propto k_z^{-0.3}$. We furthermore determined the $\sigma$ coefficient that characterises the relative importance of gradients in the turbulent velocity and density stratification, respectively. For the case of compressible ISM turbulence, this had previously found to be $\sigma\simeq \nicefrac{1}{3}$ \citep[see][]{2013ApJ...762..127B}. Here recover an even smaller value of $\sigma\simeq 0.2$, with a weak scale dependence.

In a recent attempt to measure the mean-field coefficients by means of statistical methods \citep{2020MNRAS.491.3870B}, we had noticed a certain discrepancy in the overall amplitude of the obtained effects when comparing TF results, on the one hand, with those obtained via singular value decomposition (SVD) as well as the (non-scale dependent) regression method suggested by \citet{2002GApFD..96..319B}, on the other hand. There we already conjectured that the difference may simply be a result of the intrinsic scale-dependence of the effects, and that the statistical methods are only sensitive to the (comparatively small) scales that are in fact implemented within the DNS. This notion is now broadly confirmed by our new scale-dependent TF results, which highlights that the spectral TF method provides a more complete and at the same time less biased assessment of the transport effects. We deem this more than adequate compensation for its increased computational demand.

\subsection{Memory effects} 

We have determined for the first time closure coefficients expressing the dependence of the mean EMF on the \emph{history} of the mean-magnetic field for the case of the turbulent ISM. The results for the dynamo effect, $\alpha_{\phi\phi}(z,k_z^{\rm TF}\!\! = \!2\pi/L_z,\omega^{\rm TF})$, and accordingly for the turbulent diffusion are very similar to the generic case of helically forced, mildly compressible turbulence (see \Section{benchmark_time}). In contrast to this, the profile for the diamagnetic pumping, $-\gamma_z (z,k_z^{\rm TF}\!\! = \!2\pi/L_z,\omega^{\rm TF})$ shows a potential sign change at intermediate angular frequencies, which, however, is not distinguishable from the ``response'' being fully out-of-phase with respect to the causing fluctuating mean field. Taken together, we have laid the foundation for a comprehensive mean-field modelling of the Galactic dynamo. While our results suggest that the conventional (that is, local and instantaneous) closure assumption is generally justified, the interesting behaviour seen in the pumping term warrants to study how this may affect the properties of derived dynamo models.

In retrospect, when comparing with the results involving helical forcing \citep[see][and \Section{benchmark_time}]{2009ApJ...706..712H}, sampling to even lower frequencies would have provided a more complete picture for the ISM case as well -- but the saturation effect is already clearly discernible with the current coverage. Moreover, obtaining converged results with good signal-to-noise at low oscillation frequencies also appears to require a longer simulation timeline, escalating the computational demands even further.\footnote{The two MHD simulations carried out here have consumed 144,000 core hours in total, with solving the TFs amounting to 63\% and 80\% of the total workload, for the 20 and 44 extra vector equations, respectively.} In view of the independent character of each of the TF equations, improving the density and dynamic range of the frequency-sampling may be computationally attainable via offloading this part of the workload to multi-core architectures within the MHD code in the future, even though the attainable speedup may in fact be limited by restrictions in the memory bandwidth.

Even without these technological advancements, we have demonstrated that the spectral TF method of \citet*{2008A&A...482..739B} and its time-dependent counterpart introduced by \citet{2009ApJ...706..712H} are highly useful tools and indeed produce interesting results for advanced applications, such as our investigation of realistic high-Mach-number turbulence driven by supernovae in the multi-phase interstellar medium, where mean-field effects occur naturally as an outcome of the combined action of gravity and the Coriolis force. Notwithstanding the challenges of understanding the quenching in the strong-field limit \citep*{2013MNRAS.429..967G}  and covering the parameter space required by genuinely global models \citep*{2013A&A...560A..93G}, combining these approaches \citep[as pioneered by][]{2012AN....333...71R} will allow to derive the most comprehensive mean-field closures achievable.


\section*{Acknowledgements}
We thank Axel Brandenburg, Mart{\'i}n Pessah, Abhijit Bendre and Kandaswamy Subramanian for useful discussions. This work used the \NIRVANA code version 3.3, developed by Udo Ziegler at the Leibniz-Institut f{\"u}r Astrophysik Potsdam (AIP). All computations were performed on the \texttt{Steno} node at the Danish Center for Supercomputing (DCSC).




\appendix

\section{Benchmark using helical forcing in the kinematic limit} 
\label{sec:benchmark}

We verify our implementation of the spectral TF method for the simple test case of helically-forced (mildly compressible) hydrodynamic turbulence, that is, in the strict kinematic limit of the mean-field dynamo, where the back reaction via the Lorentz force is ignored. To this end, we perform a hydrodynamic DNS in a cubic domain, solving
\begin{eqnarray}
      \partial_t\,\rho + \nabla\cdot\left(\rho \V\right) & = & 0\,,
      \nonumber\\
      \partial_t(\,\rho\V\,) + \nabla\cdot(\,\rho\mathbf{vv}\,) & = &
      - \nabla p + \nabla\cdot\tau + \bm{f}(\kk,t)\,,
\end{eqnarray}
along with the TF induction \Equation{testfields}, and using $p=c_{\rm s}^2\,\rho$ as the equation of state. We closely follow \citet{2001ApJ...550..824B} and adopt a helical forcing function $\bm{f}(\kk,t) \equiv N\,\Re\left\{\bm{f}_{\kk}\, \exp\,i\,(\kk\cdt{\rm r}+\phi)\right\}$, with $N\equiv f_0 c_{\rm s}\,(|\kk|\,c_{\rm s}/\delta t)^{1/2}$ the normalisation with respect to the numerical timestep, $\delta t$, and $f_0\simlt 1$ a dimensionless forcing amplitude. The forcing obtains its helical character via
\begin{equation}
  \bm{f}_{\kk} \equiv
  \frac{\kk\tms(\kk\times\hat{\bm{e}})
    -i\, |\kk|\,(\kk\times\hat{\bm{e}})}
       {2|\kk|^2 \sqrt{1-(\kk\cdot\hat{\bm{e}})^2/|\kk|^2}}\,,
\end{equation}
with $\hat{\bm{e}}\!\ne\!\kk$ an isotropically sampled arbitrary unit vector required for constructing a vector $(\kk\times\hat{\bm{e}})\perp\!\kk$.

Note that $|\bm{f}_{\kk}|=+1$, and --- $\bm{f}$ being an eigenfunction of the curl operator --- we have $i\,|\kk|\times \bm{f}_{\kk} = |\kk| \bm{f}_{\kk}$, such that the helicity of the forcing, $\bm{f}\cdot\nabla\times\bm{f}$, is positive definite. The wave vector $\kk(t)$ is chosen from a list of 66 (158) pre-computed discrete wave vectors obeying the periodicity of the box, and that have $|\kk| \in [\kf\!-\!\Delta k,\,\kf\!+\!\Delta k]$, with $\kf=3$ ($\kf=5$), and $\Delta_k=1/4$. Note that both $\kk(t)$ and the random phase $\phi(t)$ are updated after each timestep, $\delta t$, of the DNS.

\subsection{Non-locality in space} 
\label{sec:benchmark_space}

\begin{figure}
  \center\includegraphics[width=0.9\columnwidth]{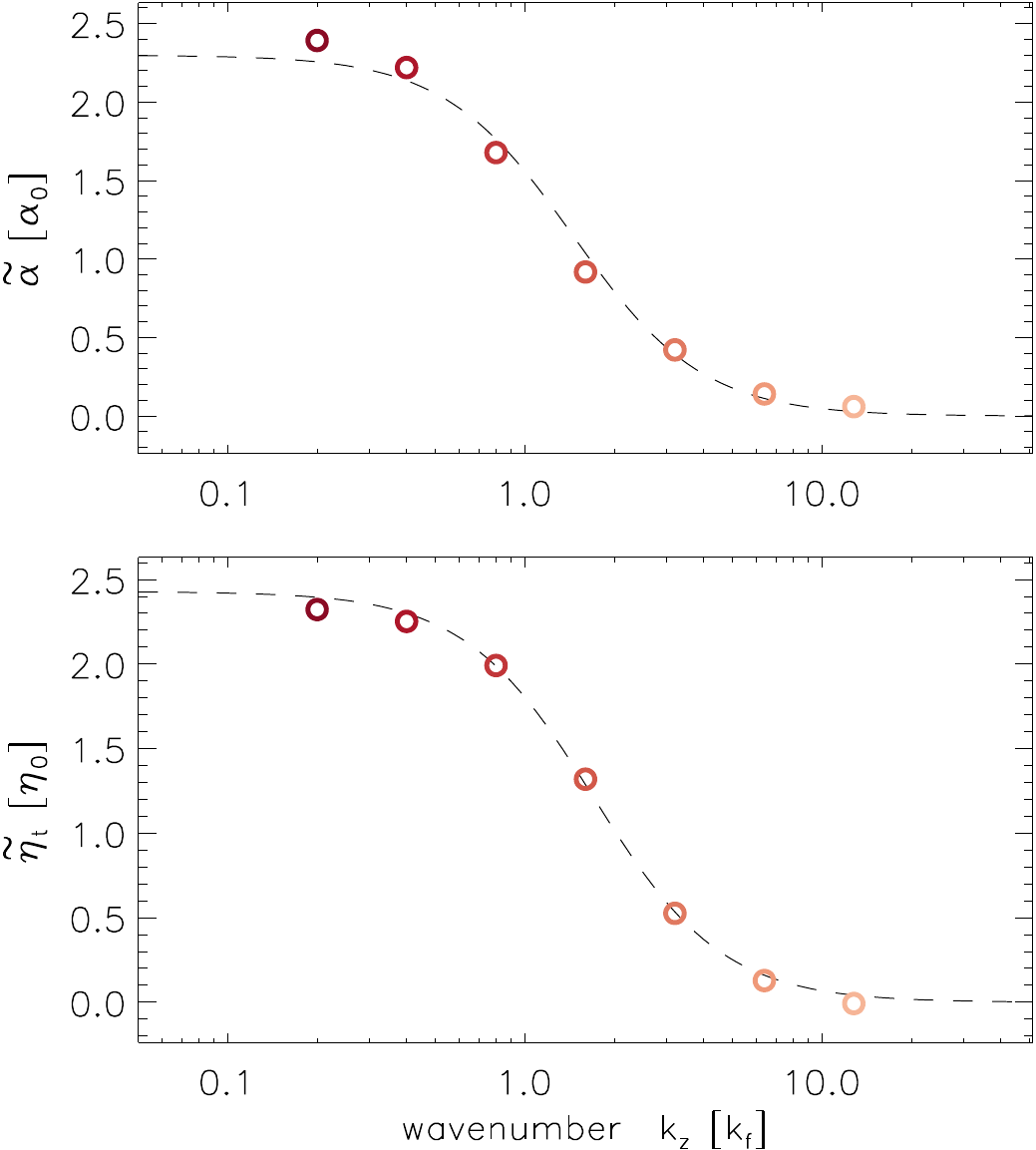}
  \caption{Spectral amplitudes for the $\alpha$~effect (top panel) and turbulent diffusion, $\etat$ (lower panel) for helically-forced isothermal turbulence with $\kf=5$, ${\rm Ma}\simeq 0.1$, and ${\rm Re}=\Rm\!\simeq\! 10$. Dashed lines indicate Lorentzian fits with $k_{\rm c}^{(\alpha)}\! = \!1.45$ and $k_{\rm c}^{(\eta)}\! = \!1.70$, in units of $\kf$, respectively.}
  \label{fig:k_helical}
\end{figure}

In \Figure{k_helical}, we show the spectral dependence of $\tilde{\alpha}(k_z)$ and $\tilde{\eta}_{\rm T}(k_z)$, adopting units of $\alpha_0 \equiv -\nicefrac{1}{3}\,v_{\rm rms}$ and $\eta_0 \equiv \nicefrac{1}{3}\,v_{\rm rms}\,\kf^{-1}$. Both curves are very well approximated by a Lorentzian with a characteristic length scale $k_{\rm c}\simeq\nicefrac{3}{2}\,\kf$. This is in contrast to the results of \citet*{2008A&A...482..739B}, who find $k_{\rm c}\!=\!\kf$ for $\tilde{\alpha}(k_z)$, and $k_{\rm c}\!=\!2\,\kf$ for $\tilde{\eta}_{\rm T}(k_z)$, respectively, for a very similar test case with $\kf=5$ and $\Rm=10$. While the qualitative trend is the same, the difference is much less pronounced in our case.

\subsection{Non-locality in time} 
\label{sec:benchmark_time}

For the purpose of fitting our obtained frequency response for the case of helically forced turbulence, we write \Equation{osc_decay} separated into its real and imaginary part, that is,
\begin{eqnarray}
  \Re & = & \frac{1 \,+\, (\omega^2+\omega_0^2)\,\tauc^2 }%
      { 4\,\omega^2\tauc^2
        + \left( 1 \,-\, (\omega^2-\omega_0^2)\,\tauc^2 \right)^2}\,,
  \label{eq:re_fit}
  \\[4pt]
  \Im & = & \frac{1 \,+\, (\omega^2-\omega_0^2)\,\tauc^2 }%
      { 4\,\omega^2\tauc^2
        + \left( 1 \,-\, (\omega^2-\omega_0^2)\,\tauc^2 \right)^2}
      \;\omega\,\tauc\,,
  \label{eq:im_fit}
\end{eqnarray}
with distinct fit parameters, $\tauc$ and $\omega_0$ for $\tilde{\alpha}(\omega)$ and $\tilde{\eta}(\omega)$, respectively, but where, for both quantities, we fit the real and imaginary parts simultaneously. Note that, because the amplitude of the effect deviates from the natural units, $\alpha_0$ and $\eta_0$, introduced in \Section{benchmark_space}, we moreover add a third parameter, $A_0\!\simeq\!1$, accounting for the overall amplitude of the effect.

\begin{figure}
  \center\includegraphics[width=0.9\columnwidth]{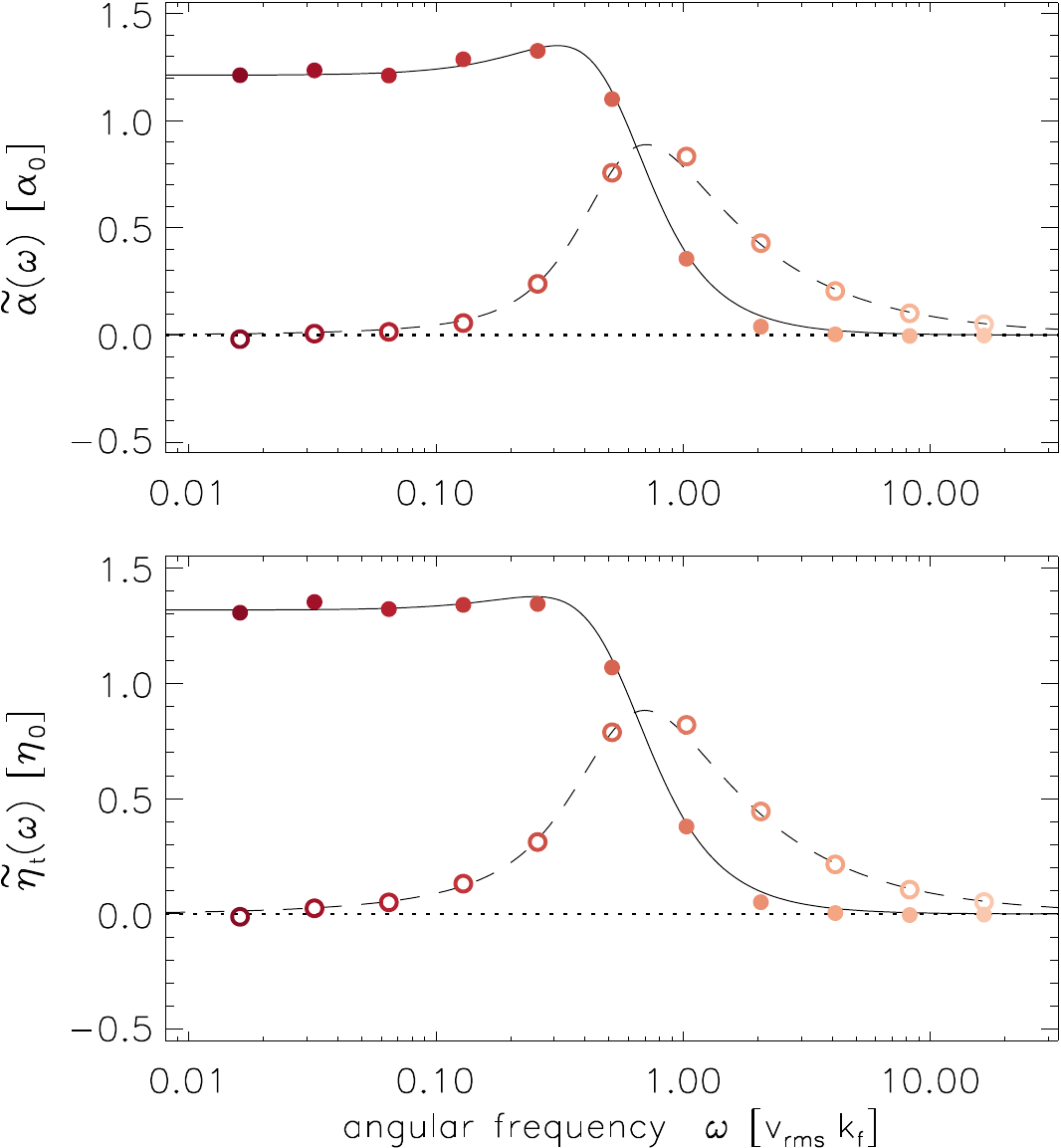}
  \caption{Frequency-dependent amplitudes of the $\alpha$~effect (top panel) and turbulent diffusion, $\etat$ (lower panel) for helically-forced isothermal turbulence with $\kf\simeq 3$, ${\rm Ma}\simeq 0.1$, and ${\rm Re}=\Rm\simeq 22$. The solid and dashed lines show a simultaneous fit to \Equation{re_fit} and (\ref{eq:im_fit}) for the real and imaginary part, respectively (see \Table{fit}, for a listing of the coefficients).}
  \label{fig:w_helical}
\end{figure}

The corresponding results are plotted in \Figure{w_helical}, that can be directly compared to figure~12 in \citet{2009ApJ...706..712H}, which shows results obtained with the sixth-order \textsc{Pencil} code. Qualitatively, the results compare rather well, but with notable discrepancies, when going into details. As expected on theoretical grounds, the transition occurs around $\omega=v_{\rm rms}\,\kf$, where the real part rises roughly to its value at low frequencies, and where the imaginary part has a broad peak around this transition, indicating a phase lag in the response. The oscillatory part of the function leads to a deviation from a purely step-like behaviour and leads to a mild peak in the real part around $\omega \simeq v_{\rm rms}\,\kf/3$. In agreement with \citet{2009ApJ...706..712H}, this peak is more pronounced in $\tilde{\alpha}(\omega)$, compared with $\tilde{\eta}(\omega)$. The effect displays itself considerably starker in their results, however.

\begin{table}
  \label{tab:fit}
  \begin{center}
  \begin{tabular}{lcccccc}\hline\hline
    & $A_0$ & $\tauc$ & $\omega_0$ & ${\rm Sr}$ & $\omega_0\,\tauc$ \\
    \hline
    $\tilde{\alpha}(\omega)$ & 2.15 & 1.25 & 0.70 & 2.44 & 0.88 \\
    $\tilde{\eta  }(\omega)$ & 2.07 & 1.17 & 0.65 & 2.27 & 0.76 \\
    \hline
  \end{tabular}
  \caption{Fit coefficients for the curves from \Equations{re_fit}{im_fit}, and derived parameters ${\rm Sr}\equiv \tauc v_{\rm rms}\,\kf\simeq 1$, and $\omega_0\,\tauc \simeq1$, with the latter two as well as $A_0$ expected to be around unity for fully-developed turbulence.}
  \end{center}
\end{table}

We list the obtained fit coefficients along with the Strouhal number, ${\rm Sr}\equiv \tauc v_{\rm rms}\,\kf$, and the product of the coherence time and oscillation frequency, $\omega_0\,\tauc$, in \Table{fit}. We generally find mild deviations from the expected magnitudes of these numbers, which may not be too discomforting in view of the moderate value of $\Rm\simeq 22$ that we assumed. The discrepancies with the results of \citet{2009ApJ...706..712H}, however, do indeed warrant further investigation.


\bsp

\label{lastpage}

\end{document}